\documentclass[twoside,letterpaper]{soups}
\usepackage[hyphens]{url}
\usepackage{xcolor}
\usepackage{subfigure}
\usepackage{algorithm}
\usepackage{tabularx}
\usepackage{fancyvrb}
\usepackage{comment}
\usepackage{balance}

\begin{document}

\newcommand{\tehila}[1]{{\color{red} [Tehila: #1]}}

\title{Leveraging Personalization To Facilitate Privacy}
%

\numberofauthors{2}
\author{
\alignauthor Tehila Minkus\\
    \affaddr{New York University}\\
    \affaddr{Brooklyn, NY}\\
    \email{tehila@nyu.edu}
\alignauthor Nasir Memon\\
    \affaddr{New York University}\\
    \affaddr{Brooklyn, NY}\\
    \email{memon@poly.edu}
}

\maketitle

\begin{abstract}
Online social networks have enabled new methods and modalities of collaboration and sharing. These advances bring privacy concerns: online social data is more accessible and persistent and simultaneously less contextualized than traditional social interactions. To allay these concerns, many web services allow users to configure their privacy settings based on a set of multiple-choice questions.

We suggest a new paradigm for privacy options. Instead of suggesting the same defaults to each user, services can leverage knowledge of users' traits to recommend a machine-learned prediction of their privacy preferences for Facebook. As a case study, we build and evaluate MyPrivacy, a publicly available web application that suggests personalized privacy settings. An evaluation with 199 users shows that users find the suggestions to be appropriate and private; furthermore, they express intent to implement the recommendations made by MyPrivacy. This supports the proposal to put personalization to work in online communities to promote privacy and security.
\end{abstract} 





\section{Introduction}
\label{intro}
If the twentieth century was the era of mass production, the twenty-first has become the era of personalization. Particularly in the online arena, services have become intensely focused on the individual. The user is encouraged to create content, and the content he consumes is tailored specifically to his interests. New tracking technologies enable companies to build extensive dossiers on a user's habits and hobbies, and these in turn allow them to serve him ads, articles, and videos that best conform to his interests.

However, one common domain of online services remains devoid of any automatic individualization: privacy controls. While privacy options are offered by the leading social networks, they are presented in a ``one-size-fits-all'' mold: each user's profile is initialized to the same default settings, regardless of his unique traits. Moreover, the interfaces are often difficult to navigate. Users are notoriously bad at access control decisions, and the privacy interfaces offer little help in the way of feedback.
In Facebook, for example, users are given 18 settings, each with multiple choices. Users are not able to easily deduce which configuration best fits their needs and goals, and the current defaults are generally incompatible with users' true privacy preferences.

Particularly in the setting of online collaborative and social environments, this can degrade a user's experience and the service's utility. When privacy options are confusing, users' final choices are not always in line with their actual preferences. This can serve as a barrier to engagement in social networks \cite{nov2009social}, and it can also lead to harmful side effects when users share information with unseen audiences \cite{wang2011regretted}. Since trust is essential to building and maintaining healthy online communities, manageable privacy is a key aspect of any social online service, as Malhotra et al. point out \cite{malhotra2004internet}.

Personalization has been offered as a panacea to the information overload threatened by the proliferation of web content. In this paper, we show that it can also be used as a privacy-enhancing tool to facilitate trust in online communities and services. We choose Facebook privacy settings as a case study to demonstrate how personalization can be leveraged for privacy-positive purposes online. Via an online survey ($n=451$), we examine the privacy settings of users on Facebook and relate their choices to demographic and personality features. Subsequently, we use these outcomes to build and deploy an online application, MyPrivacy, that automatically recommends privacy settings based on a user's demographic and personality traits. MyPrivacy compares a user to other similar users, using this as a basis to derive suggestions for privacy settings. A second online user study ($n=199$) shows that users find MyPrivacy's recommendations to be more appropriate and private than the most popular settings chosen by users. This points to a clear demand for a more personalized approach to online privacy.

Our main contributions are:
\begin{itemize}
\item Proposing personalization as a method for guided privacy configuration in online services and communities.
\item Designing and building MyPrivacy, a case study of a personalized recommender system that incorporates relevant personal traits into users' privacy settings.
\item Evaluating MyPrivacy via a randomized user study of 199 users to show increased relevance, privacy, and usability in the personalized model of privacy settings as compared to a baseline.
\end{itemize}

The structure of this paper is as follows: first, we present an overview of related research in the literature. Next, we conduct a case study of Facebook privacy settings to demonstrate the implementation of personalization as a tool towards better privacy management in online services and communities. We show how users' privacy preferences on Facebook are related to their personal and social traits. Based on this, we propose and implement MyPrivacy, a system for learning and recommending privacy settings to individual users. We evaluate MyPrivacy against a baseline model in a randomized test and show that it outperforms it in perceived accuracy and privacy and in the users' intended adoption of the recommendations. Consequently, we discuss the import and some takeaways of our findings and enumerate some directions for further work. Finally, we summarize and conclude.

\section{Background and Related Work}
\label{sec:related}

\subsection{Personalization}
Personalization, or the automatic tailoring of an application's content or behavior to an individual's traits, has been examined at length in the human-computer interaction community in many applications, such as vehicles \cite{sadler2004vehicle}, robotics \cite{sung2009pimp}, and even public media streaming \cite{mahato2008implicit}. Chellappa and Sin discussed the tension between privacy and the surveillance demanded by many personalization applications \cite{chellappa2005personalization}. However, we are aware of no prior work that uses personalization  (i.e. machine learning methods) to automatically infer a user's privacy or security preferences in local, online, or social applications.

\subsection{Online Social Network Usage and Personality}
There is a good deal of research providing basis for the idea that usage of online social networking will vary with a user's personality and traits. For example, Guercia et al.~\cite{quercia2012personality} examine the personality traits that predict Facebook popularity. They find that the number of a user's Facebook friends is correlated with her degree of extraversion and inversely correlated with her degree of neuroticism.

Similarly, Ross et al.~published a study~\cite{ross2009personality} of Facebook usage and motivations based on the Five Factor Model of personality. Subjects scoring high on openness were found to be more social online; however, other similar hypotheses regarding extraversion and agreeableness were not upheld. Christofedes et al.~\cite{christofides2009information}
research the connection between information control and disclosure
on Facebook and whether they are predicted by the same personality factors. Their findings  showed that people with a greater need for popularity are more likely to disclose information on Facebook, whereas people with higher levels of trust and self-esteem exhibit stricter information control.

\subsection{Privacy and Personal and Social Traits}
Moreover, several studies have shown a clear link between users' privacy orientations and their personal traits, such as personality, sociability, and demographics. Junglas et al.~\cite{junglas2008personality} researched the connection between personality and privacy perception in
the context of location-based services. Out of the five personality traits specified by the
Five Factor Model, they found that subjects scoring high on agreeableness, conscientiousness,
 and openness to experience expressed fewer privacy concerns about location-based services. Halevi et al. \cite{halevi2013pilot}
 found that users who were more open were also less likely to be private on Facebook.

Research has shown correlations between personality and private behavior in the offline context as well.
In 1982, Pederson~\cite{pedersen1982personality} showed that subjects low in self-esteem
were more likely to seek solitude and anonymity. Those with low esteem for others were more prone to seek solitude and intimate situations, whereas happy-go-lucky people tended away from anonymity. Introspective individuals were likely to be reserved but less likely to choose isolation and intimacy with family. Finally, tolerant people tended to prefer anonymity. This work focuses on non-digital interactions with different metrics than our research, but it provides evidence for the hypothesis that personality is linked with private behavior.

Social factors were also found to correlate with privacy settings on Facebook by Lewis et al. \cite{lewis2008taste}, who found that college-aged Facebook users were more likely to be private if their friends or roommates had more private settings.

The connection between demographic factors and private behavior in online social networks has also been examined
in the literature. Christofedes et al.~\cite{christofides2012hey} researched whether
age impacts private behavior on Facebook. Despite popular sentiment to the contrary, their
findings show that adults and adolescents generally exhibited very similar behavior on
Facebook. Dey et al.~\cite{dey2012privacy} showed that users in more affluent neighborhoods
in NYC had more private Facebook settings. Gilbert et al., in a 2008 study about MySpace,
found that rural users had more private profiles and more exclusive friends networks
~\cite{gilbert2008network}.

boyd and Hargittai \cite{boyd2010facebook} found that the most salient factors related to making modifications in privacy settings were users' Facebook habits and their general level of Internet skill. Similarly, Lewis et al. \cite{lewis2008taste} found that among college students, frequent use of Facebook predicted more private settings.

Fogel and Nehmad investigated gender differences in privacy
and found that, although women expressed more privacy concerns than men, their self-reported
behavior on Facebook was no more private than that of men~\cite{fogel2009internet}. Hoy and Milne \cite{hoy2010gender} found in a 2010 survey of Facebook users aged 18-24 that women reported more concern for privacy and were also more likely to engage in proactive privacy-preserving behavior, such as reading privacy policies and changing their privacy settings. Lewis et al. \cite{lewis2008taste} also found that female college students had more private settings on Facebook. However, boyd and Hargittai \cite{boyd2010facebook} did not find strong gender differences among young adults' approach to privacy settings in their 2010 research.

A Pew Research poll of teens in 2013 \cite{pew2013teens} showed ethnic differences in social media privacy behaviors. For example, African-American teens were considerably more likely to use pseudonyms on social media than white teens (23\% versus 5\%). Overall, 39\% of African-American teens surveyed reported posting fake information to their profiles, compared with 21\% of white teens.

\subsection{Simplifying Privacy}
 Privacy settings are notoriously difficult to navigate: Liu et al.~\cite{liu2011analyzing} found that 67\% of the time, Facebook users expressed sharing preferences for photos that were different than their actual settings. In response, existing research has discussed recommender systems as a way to make privacy
more accessible to the average user. Bonneau et al.~\cite{bonneau2009privacy} propose an
expert-based system, termed privacy suites. Privacy suites are predefined configurations,
chosen either by experts or members of users' social networks, which are presented to
the user as an alternative to the less-private automatic defaults. In a somewhat different
approach, Watson et al.~\cite{watson2009configuring} attempt to help users attain privacy more easily by presenting the
privacy settings in a more visual layout. They propose AudienceView, an alternative privacy
policy interface to provide a better mental model and visual feedback for privacy
management on social network sites.

Fang and LeFevre~\cite{fang2010privacy} build a system to semi-automatically assign users'
Facebook friends to lists with different privacy settings. Using active machine learning techniques,
they allow the user to specify settings for a certain number of friends and then extend those
settings to other friends with similar features. Also,  Stutzman and Kramer-Duffield~\cite{stutzman2010friends}
explore an alternative way of enhancing privacy by maintaining a social network site profile
to be friends-only.

In a similar vein, Ravichandran et al. \cite{ravichandran2009capturing} attempt to learn several default policies for location sharing by studying the sharing preferences expressed by users in realtime. They use a binary classification for each sharing opportunity, and then they build three generalized default configurations based on their data in an attempt to balance both expressiveness and ease for privacy settings. However, Ravichandran et al. do not attempt to personalize the results; they aim to present three reasonable choices as defaults for a user who is configuring his profile.

\subsection{Significance of This Paper}
While some approaches have been proposed towards easing the difficulty of choosing privacy settings in online communities, none have leveraged personalization as a tool towards better defaults. We build upon existing literature showing links between personal traits and privacy preferences to offer a usable framework for configuring personalized privacy settings. To demonstrate the potential of personalization as a privacy tool, we propose and implement a case study based on the Facebook privacy settings. Our proposed system, MyPrivacy, allows Facebook users to avoid hard access-control questions by having them answer easy questions about themselves; these facts are then incorporated into a machine-learned recommender for privacy settings. This approach can also be applied to privacy settings in other social and mobile applications.

\section{Modeling Privacy}
\label{sec:hypotheses}

Personalization can be utilized to help automate or guide privacy and security in online communities and services. To illustrate the idea, we examine the popular online social network of Facebook. In order to model users' privacy choices, we examine three possible areas: personality traits, demographic factors, and self-reported privacy concerns.

\subsection{Personality Traits}

We hypothesize that personality types correlate with Facebook privacy settings (i.e., some personality types are more likely to choose more private settings on Facebook).
We use the Five-Factor Model as a system to quantify personalities. The Five-Factor Model (FFM), elaborated by McCrae et
al. \cite{mccrae1992introduction}, considers the five personality traits of
neuroticism, extroversion, openness, agreeableness, and conscientiousness. It has been
used as a personality model in several works in the literature, including Ross et al.~\cite{ross2009personality} and Halevi et al \cite{halevi2013pilot}.
The personality traits comprising the FFM are defined as follows:
\begin{itemize}
    \item Openness to new experiences: open individuals are artistic, curious, imaginative, insightful, original, and have wide interests.
    \item Conscientiousness: conscientious people are efficient, organized, planful, reliable, responsible, and thorough.
    \item Extraversion: extraverted individuals are active, assertive, energetic, enthusiastic, outgoing, and talkative.
    \item Agreeableness: agreeable people are appreciative, forgiving, generous, kind, sympathetic, and trusting.
    \item Neuroticism: neurotic people are anxious, self-pitying, tense, touchy, unstable, and worrying.
\end{itemize}

\subsection{Demographic Information}

We hypothesize as well that a user's demographic information has an impact on his privacy settings. Towards this end, we analyze the following demographic factors: gender, age, relationship status, and ethnicity.

\subsection{Privacy Concerns}
Additionally, we analyze the relationship between a user's self-reported concern for online privacy and his actual behavior, as evinced by privacy settings. Are privacy-aware users more likely to pick private settings. or does this area exhibit the common dichotomy between intentions and behavior? 
\section{Survey Methodology}
\label{sec:method}

In our case study of Facebook privacy settings, we surveyed 522 users in September 2013 to collect relevant personal data. 
The survey was hosted on SurveyMonkey.com, which provides services for creating surveys, designing the necessary charts and graphs, and efficiently analyzing the results.
\subsection{Survey Design}

The survey consisted of three main sections. In the first section, we asked the participants for demographic information
in order to draw correlations between these and Facebook privacy settings. The second section of the survey was a shortened version of the FFM personality test, called the Mini-IPIP \cite{donnellan2006mini}. In the last part, participants were asked to report their privacy settings. 

Research plans for this study were submitted to the university Institutional Review Board (IRB) and found to be exempt. Subjects were notified that the survey was being conducted for research purposes and that their data would be stored anonymously. The only mention of intent for the survey was that it was studying personality and Facebook use.

In order to ensure that our data analysis would only include attentive responses (rather than subjects who had merely clicked at random in order to earn the incentive), we incorporated attention-measuring questions into our survey. These questions were designed to look like the rest of the questions but included directives to select a specific answer. If a user did not choose the indicated option, we assumed that he was not paying attention to the survey; such users were eliminated from the final analysis.

\subsubsection{Demographic Information}

The survey's first section collected demographic information. Each subject supplied her age, gender, ethnicity, and relationship status by selecting from a list of best-fit options.

\subsubsection{The Big Five Personality Test}
\label{sec:FFM}
The next section of the survey was the personality test,
which was based on the Five-Factor Model.
In order to reduce the length of the survey, we used the Mini-IPIP version of the test \cite{donnellan2006mini}. This includes 20 questions, comprised of four questions for each of the five traits.
Each question ascribes a tendency to the subject, and subjects selected options
from a 5-point Likert scale ranging from ``Strongly Disagree" to ``Strongly Agree".

\subsubsection{Facebook Privacy Settings}

After the personality test, subjects were asked if they had Facebook accounts.
Subjects with Facebook accounts were directed to
open their Facebook privacy settings via precise and exact directions to the interface.
They were told to select their current privacy settings while consulting the Facebook privacy interface.
The conclusion of the survey asked users some related questions,
such as how satisfied they are with their current settings.
The final question asked subjects to rate their degree of concern about their online privacy.

\subsection{Recruitment}
The survey was posted as a task on Amazon Mechanical Turk,
a crowdsourcing venue that allows access to a wide range of subjects and viewpoints.
The task was restricted to US workers. Workers were compensated \$0.50 for their time.

\subsection{Participants}
Of the 522 respondents, 451 completed the survey in its entirety and responded correctly to the attention-measuring questions.
As Table~\ref{tab-demo} shows, respondents were
diverse with regards to age, gender and relationship status. Note that we restricted
our survey to adults and do not include users below
18 years of age. We also only accepted responses from within the US since the effects of culture on privacy are beyond the scope of this study and have already been analyzed in the literature; for example, see the work by Kumaraguru and Cranor regarding privacy mores in India \cite{kumaraguru2006privacy} or Zhang et al.'s research on differing privacy attitudes in the United States and China \cite{zhang2002characteristics}.

A comparison of the demographic statistics between our sample and the
Facebook statistics from December 2013, as reported by the Google Display Planner tool\footnote{\url{https://adwords.google.com/da/DisplayPlanner/Home}},
is shown in Table~\ref{tab-demo} (where available).

\begin{table}
\centering
\begin{tabularx}{\linewidth}{|X | X X|} \hline
&Survey Sample&Facebook\\ \hline \hline
\textbf{Age} & & \\
18 - 24 & 27.3\% & 25.3\% \\
25 - 34 & 50.1\% & 23.9\% \\
34 - 44 & 14.4\% & 20.1\% \\
45 - 54 & 4.7\% & 13.4\% \\
55 - 64 & 3.1\% & 10.4\% \\
65+ & 0.4\% & 6\% \\ \hline
\textbf{Gender} & & \\
Male & 62.3\% & 54.4\% \\
Female & 37.7\% & 45.6\% \\ \hline
\end{tabularx}
\caption{Description of our sample's demographics as compared to the demographics reported by the Google Display Planner tool in December 2013.
This table does not include the 32\% of users for whom the Google Display Planner could not determine a gender or age.}
\label{tab-demo}
\end{table}

\subsection{Limitations}
\subsubsection{Sample Population}
As shown in Table \ref{tab-demo}, our sample is not perfectly representative of the Facebook population; in particular, it overrepresents males and younger people. 

\subsubsection{Accurate Reporting}
While we did encourage our subjects to check their Facebook settings for accuracy before responding, we did not enforce this via any technical means. However, since we intended to discover a user's preferred privacy choices, this was sufficient for our purposes.
\section{Results}
\label{sec:results}

In this section we present the findings of our study relating Facebook settings to users' personal and social traits.

\subsection{Coding Attributes}
Here, we explain how we coded the different attributes in order to prepare them for statistical testing.

\subsubsection{Personality}
As explained in Section \ref{sec:hypotheses}, the Five Factor Model of personality assigns numbers for each trait based on a series of questions graded on a Likert scale. Each personality trait receives an integer score; a higher score demonstrates that the user's personality contains a high degree of the given trait.

\subsubsection{Gender, Ethnicity, Marital Status}
We coded gender as a binary variable, where 1=female and 0=male. Likewise, we coded each ethnicity as a binary variable, so that a each user's record would have several ethnicity variables but only one choice would be coded with a 1. Marital status was coded in the same way, with each user's marital status coded as 0 for all non-applicable states and as 1 for the chosen state.

\subsubsection{Age}
Age was coded as a discrete numerical field, represented by the landmark decade of each group. For example, we coded ages 18-24 as 20, ages 25-34 as 30, and so on.

\subsubsection{Privacy Concern and Satisfaction}
These questions were answered on a 5-point Likert scale, which we used to code the answers on a range from 0 to 4.

\subsubsection{Facebook Privacy Scoring}
In order to calculate numerical relationships between privacy settings and metrics for personality and demographics, we assigned privacy scores to the Facebook privacy choices that users reported in the survey. We followed the weight-based approach introduced by Minkus and Memon \cite{minkus2014scoring}. Briefly, this approach expresses the relative importance of the various privacy settings by incorporating weights, as judged by a user survey, into the total score. Each setting's choice is given a privacy grade on a scale from 0 to 1, and this is then multiplied by the weight; finally, the individual scores are summed onto a total score in the range of 0 to 10. A score of 0 represents perfect non-privacy, and a score of 10 reflects perfect privacy (with reference to the available options).
The most private account possible would score 10 points. However, none of our subjects attained this score, and we suspect
that enforcing such strict access control over a Facebook page would be contrary to the goals of social networking.

\begin{figure}
\includegraphics[width=\columnwidth]{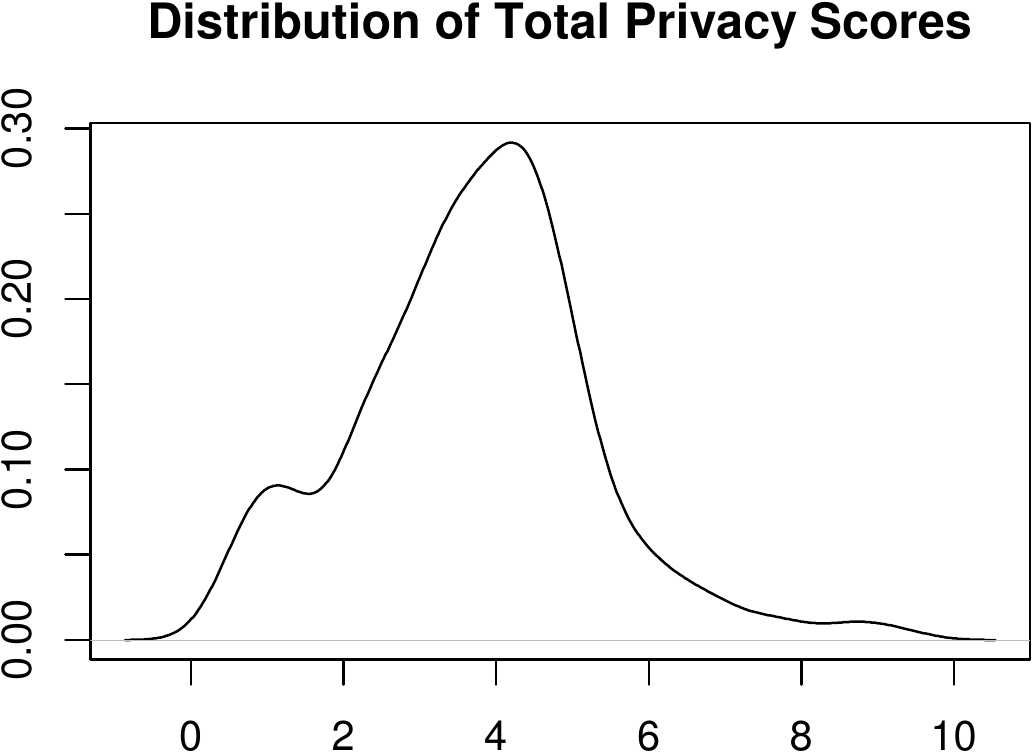}
\caption{The distribution of our sample by privacy score. The mean was 3.72 and the standard deviation was 1.58. The median score was 3.79.}
\label{fig:privacydist}
\end{figure}

The distribution of privacy scores among our subjects can be found in Figure \ref{fig:privacydist}.
The average privacy score in our pool of responses is 3.72, with a standard deviation of 1.58 and a median of 3.79. Our data shows that the vast majority of Facebook users do change their default privacy settings.
\subsection{Evaluation of Hypotheses}
To evaluate our hypotheses, we calculated the Pearson's product-moment correlation (PPMC) and $p$-values
for each trait with the total privacy scores.
The results can be found in Table \ref{tab-perscorr}.

\subsubsection{Personality and Privacy}
For our five hypotheses regarding personality, only neuroticism was found to be correlated at statistically significant levels.
The data showed that neurotic people are more likely to have private Facebook settings.
The findings regarding the other four personality traits (openness, conscientiousness, extraversion,
and agreeableness) were not statistically significant in our dataset, so we could draw no conclusions regarding their relationship with privacy settings.

\begin{table}
\label{fig:tblperscorr}
\begin{center}
  \begin{tabularx}{\linewidth}{| l  X  X |}
    \hline
    Trait & PPMC & $p$-Value \\ \hline \hline
    \textbf{Personality} & & \\ 
    Openness & -0.04 & 0.43 \\
    Conscientiousness & 0.00 & 0.98 \\
    Extraversion & -0.02 & 0.62 \\ 
    Agreeableness & 0.02 & 0.61 \\
    Neuroticism & 0.09 & 0.04* \\ \hline
    \textbf{Demographics} & & \\
    Age & -0.10 & 0.04* \\
    White & -0.09 & 0.04* \\
    Asian & 0.13 & 0.01** \\ \hline
    Concern for privacy & 0.27 & 5.09E-09** \\ \hline
  \end{tabularx}
\end{center} 
\caption{The correlations between each personality trait and privacy score.
An asterisk denotes statistical significance at $p \le .05$ level, and two asterisks denote significance at $p \le .01$ level.}
\label{tab-perscorr}
\end{table}

\subsubsection{Demographic Correlations}

\begin{figure}[t!]
\includegraphics[width=\linewidth]{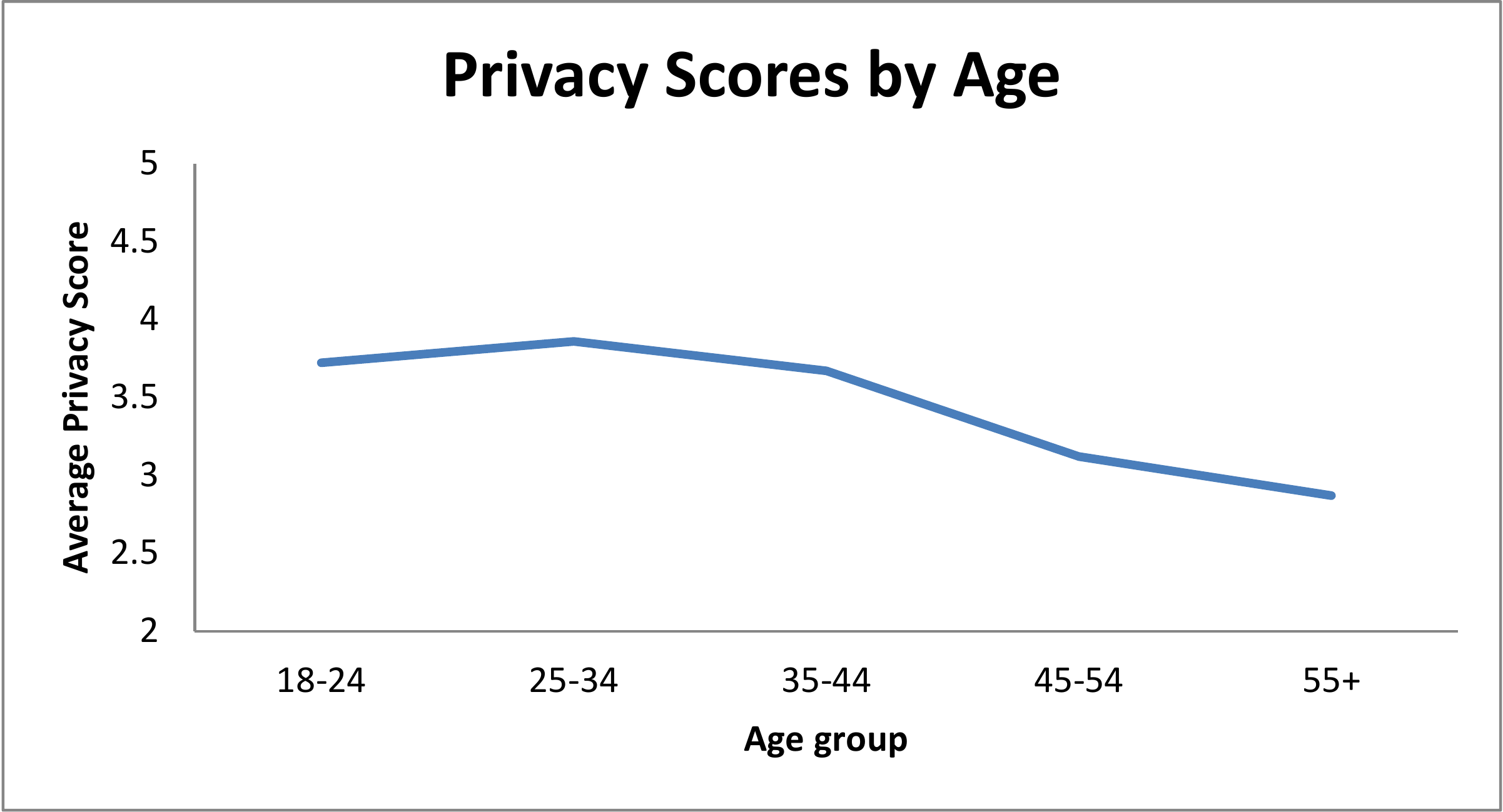}
\caption{The average privacy score for each age group. Older users are less private on average. Since the oldest group (ages 65+) had just two respondents, we combined it with the age group 55-64 for this chart.}
\label{fig:age-private}
\end{figure}

Of the demographic factors surveyed, we found that three factors were correlated
at a statistically significant level ($p \le .05$). These are shown in Table \ref{tab-perscorr}. Asians were found to be more private, while users identifying as white were found to be less private. Additionally, older users were found to be less private; see Figure \ref{fig:age-private} for a graph of each age group's average privacy score. This may be a result of lower technological fluency, as shown by boyd and Hargittai \cite{boyd2010facebook}.

\subsubsection{Concern for Privacy}
\begin{figure}[t!]
\includegraphics[width=\linewidth]{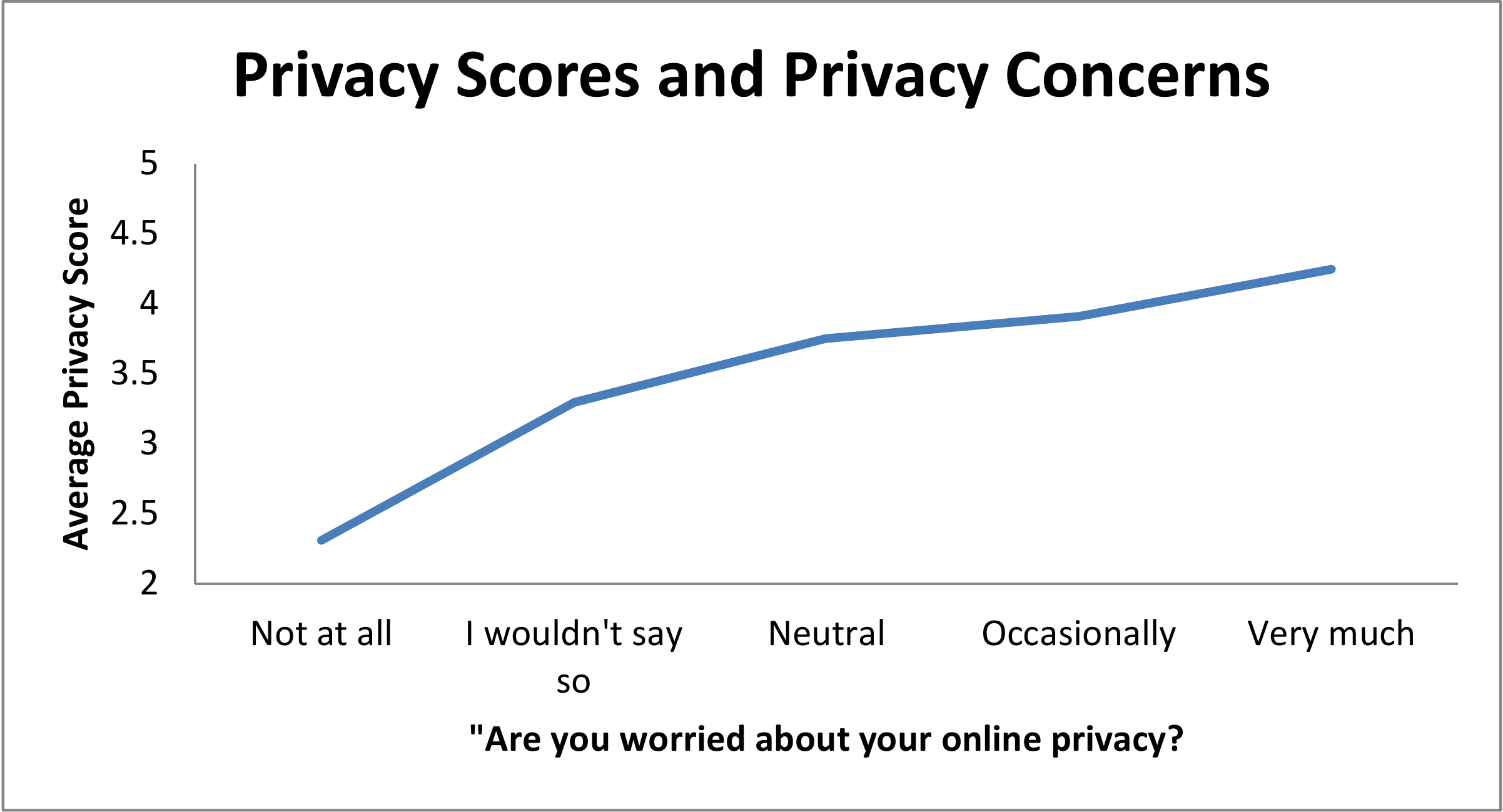}
\caption{The average privacy score for each degree of self-reported privacy concerns. As users' privacy concerns rise, so does the average privacy of their Facebook settings.}
\label{fig:concern-private}
\end{figure}

Based on our data, users who reported higher concerns for privacy followed through by selecting more private settings, as shown in Table \ref{tab-perscorr}. Figure \ref{fig:concern-private} shows how the average privacy score rises as users report more frequent privacy concerns.

\section{MyPrivacy}
\label{sec:myprivacy}
Online services are designed to be fast and easy to use. In contrast, choosing privacy settings can be very laborious and unintuitive. As a remedy, we propose a framework for privacy settings centered on an automated guide. This guide uses some simple personal information to provide a user with recommendations as a starting point for the privacy configuration process.

Our idea was inspired by the paradigm of online investment sites,
such as TradeKing\footnote{https://www.tradeking.com/}. Instead of sending the user directly to the overwhelming
list of all the trading options available, the service first collects some basic data about the user:
income, profession, assets, financial goals and risk attitude. The service then uses this information
to narrow down the options presented to the user based on the historic data of people similar to them.
This makes decisions simpler for people: instead of answering a highly specific question about investing
in a certain stock, they can answer easier, more general questions about their own life and proclivities.

This is an idea that can be applied to a diverse array of services, such as browser preferences, image sharing, and mobile phone permissions. As a case study, we build a Facebook-centric application conforming to this paradigm of privacy-positive personalization. Since privacy configurations are known to be hard for the average Facebook user~\cite{acquisti2006imagined}, we leverage the above model of privacy settings to implement a heuristic system for
recommending Facebook privacy settings. In the MyPrivacy application, we attempt to map privacy settings onto a more familiar dimension. Instead of answering
more confusing questions such as ``Do you want to share your birthdate with others' apps?", users answer
easy questions about their demographics and personality. This data is then compared against our labeled data
from Facebook users who are satisfied with their settings to generate a personalized set of privacy preferences.
A user study with $n=199$ shows overall satisfaction with the recommendations and ease of use. 
MyPrivacy shows significant improvement over static recommendation of the most popular choices for Facebook privacy settings.

These suggested privacy settings could be offered as personalized defaults for new users,
updating the outdated ``one size fits all'' standard of defaults on Facebook. This would impact different users in different ways: for privacy novices, MyPrivacy would secure them a more appropriate level of privacy than the one-size-fits-all defaults. For more experienced users, it would provide them with better context when tweaking their settings, since they would have exposure to some idea of what other users had preferred.

\subsection{System Description}

MyPrivacy utilizes our above findings to recommend privacy settings for users based on the correlations
of different personality traits and demographic factors.
We base the recommender system on the k-nearest neighbor
(kNN) machine learning algorithm, which assigns labels by finding the input's closest matches in the training data.
After running simulations for various values of $k$ in the WEKA machine learning package~\cite{hall2009weka},
we set $k=18$, so that the user's privacy setting
recommendations were generated from the average of the 18 closest matches in our database (see
Algorithm \ref{alg:recommender}).

\begin{algorithm}
\caption{Privacy Settings Recommendation}
\label{alg:recommender}
\begin{verbatim}

On inputs InputTuple, Database:
    1. Filter the Database to exclude users who
       are dissatisfied with their current settings.
    1. Search for InputTuple's 18 closest
       matches in Database
    2. For each privacy setting in InputTuple:
        a. Take the average of the setting of
           the 18 closest matches
        b. Round it to the nearest integer
        c. Set the InputTuple's setting to
           the rounded average
    3. Output the completed InputTuple
\end{verbatim}
\end{algorithm}

As noted in Algorithm \ref{alg:recommender}, we filter our initial survey data to exclude users who were dissatisfied with their current settings. Since the user's output is based on the data of users similar to himself, we wish to avoid recommending configurations that users regret choosing. After removing the users who reported themselves as highly dissatisfied, our reference database consisted of 382 records (84.5\% of the initial dataset).

\subsection{Implementation}
\begin{figure}
\fbox{\includegraphics[width=\linewidth]{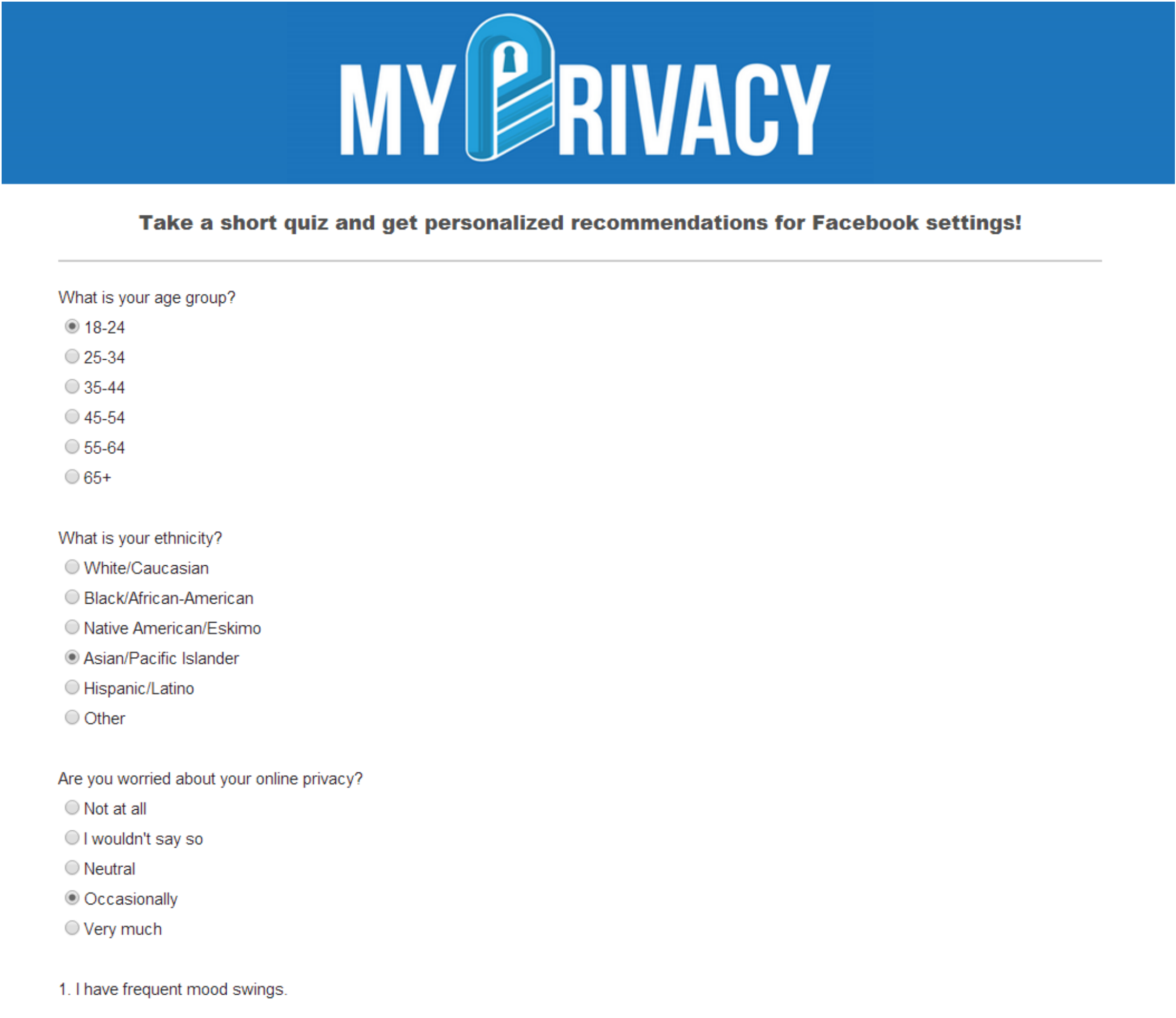}}
\caption{Screenshot of the MyPrivacy input form.}
\label{fig:screenshot}
\end{figure}

MyPrivacy is a web application implemented in HTML and PHP.
Readers can access it here\footnote{The application is hosted and linked anonymously for the duration of the review period.}: \url{http://tinyurl.com/myprivacyapp}.
For this prototype implementation of our application, we used
seven questions in total as input to the system, testing for the following traits:
\begin{itemize}
\item Age group (1 multiple choice question)
\item Ethnicity (1 multiple choice question)
\item Privacy concern (1 question on a Likert scale)
\item Neuroticism (4 questions on a Likert scale, from the Mini-IPIP test)
\end{itemize}
In other words,
we included all significant factors from our data, as shown in Section~\ref{sec:results}. A screenshot of the input page can be seen in Figure \ref{fig:screenshot}.

\begin{figure}
\fbox{\includegraphics[width=\linewidth]{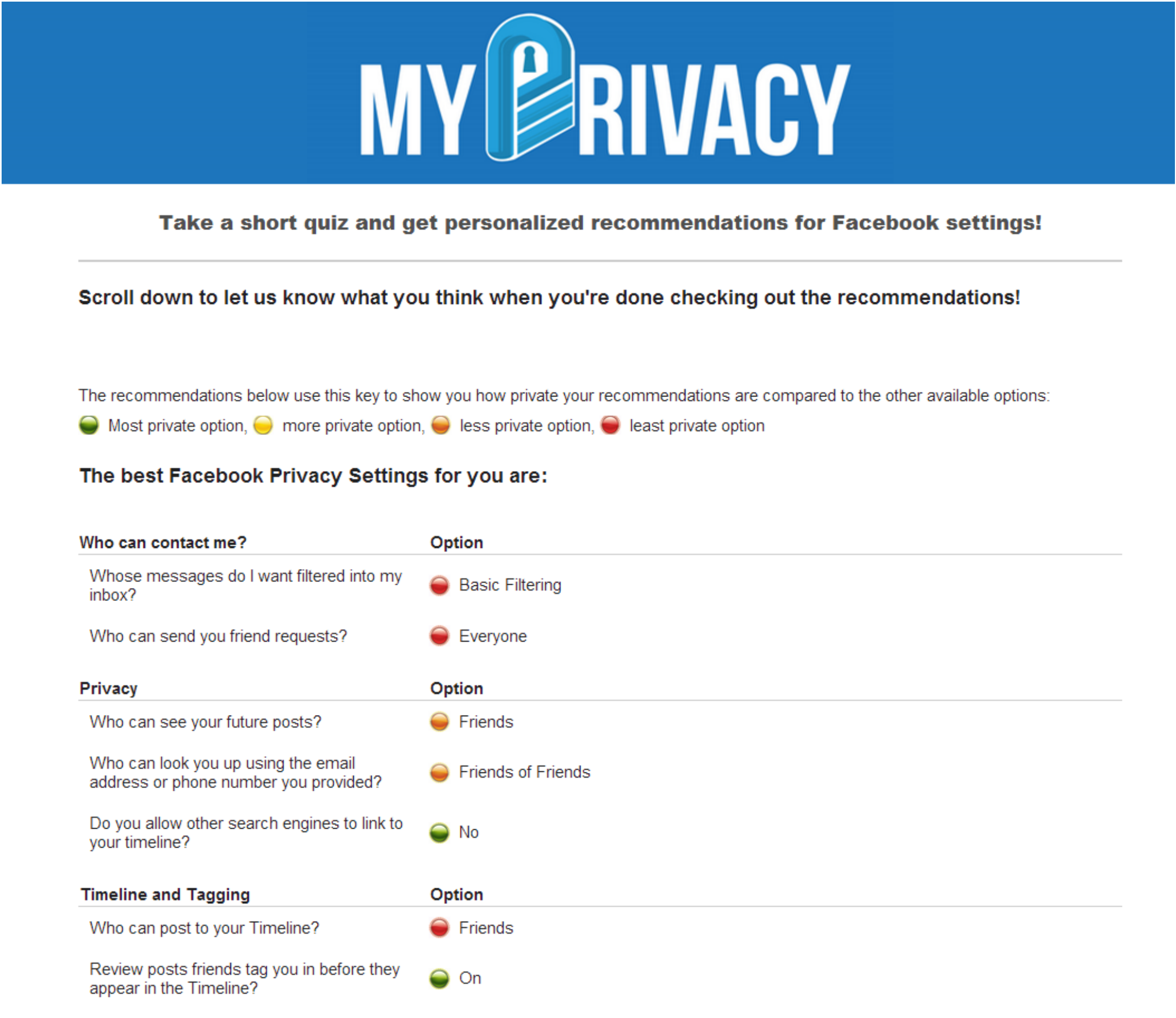}}
\caption{Screenshot of sample MyPrivacy recommendations.}
\label{fig:screenshot2}
\end{figure}

The output of MyPrivacy consists of the recommended Facebook privacy settings. 
In order to allow users to judge the privacy of the recommended settings, we utilize a color-coded key to indicate the level of privacy for each suggested option. Green is used to indicate the most private option, and red indicates the least private option; yellow and orange are used as intermediate colors. This supplies users with some context when evaluating the suggested options, since they can look at the colors and decide whether they align with their privacy preferences. A screenshot of some sample suggestions is shown in Figure~\ref{fig:screenshot2}.

\section{Evaluation}
\label{sec:eval}

In order to measure the utility and accuracy of the MyPrivacy application, we conducted a user study. In the course of the user study, we compared MyPrivacy to a baseline model in an A/B test and asked users to provide their judgments of the recommended settings. Our results showed that users preferred MyPrivacy's recommendations. In each of the systems, users viewed identical interfaces; however, the suggestions presented in the output screen were generated using different methods.

As a baseline, we tested an alternate model termed the popular test mode. In the popular test mode, users received a static recommendation page regardless of their input. The recommendations in this case were based on the most popular settings chosen in our survey. For each setting, we suggested the option that a majority (or plurality) of users had chosen in our study. This baseline was intended to see whether personalized privacy settings actually provide a better fit than simply using the most popular options.

\subsection{Metrics}
After users input their data and were shown their recommended settings, we directed them to evaluate the system. Users voiced their opinions about the application via
several questions, graded on a 5-point Likert scale:
\begin{itemize}
\item How appropriate are these recommendations for you?
\item How private are these settings?
\item Are you planning to use these settings?
\item Would you prefer to use this tool to choose your privacy settings?
\end{itemize}
There was also an open comment field available for the users to express any other opinions.

\subsection{Results}
Once again, we recruited workers from Mechanical Turk.
MTurk workers were debriefed via the main page of the HIT and paid \$0.30 for their
contributions. We provided MTurk workers with a special link to MyPrivacy, where they were randomly redirected to one of the two modes.
In total, we received 199 responses: 97 for the popular mode,
and 102 for MyPrivacy.

\begin{figure*}
\centering
 \subfigure[Appropriateness of settings]{\includegraphics[width=\columnwidth]{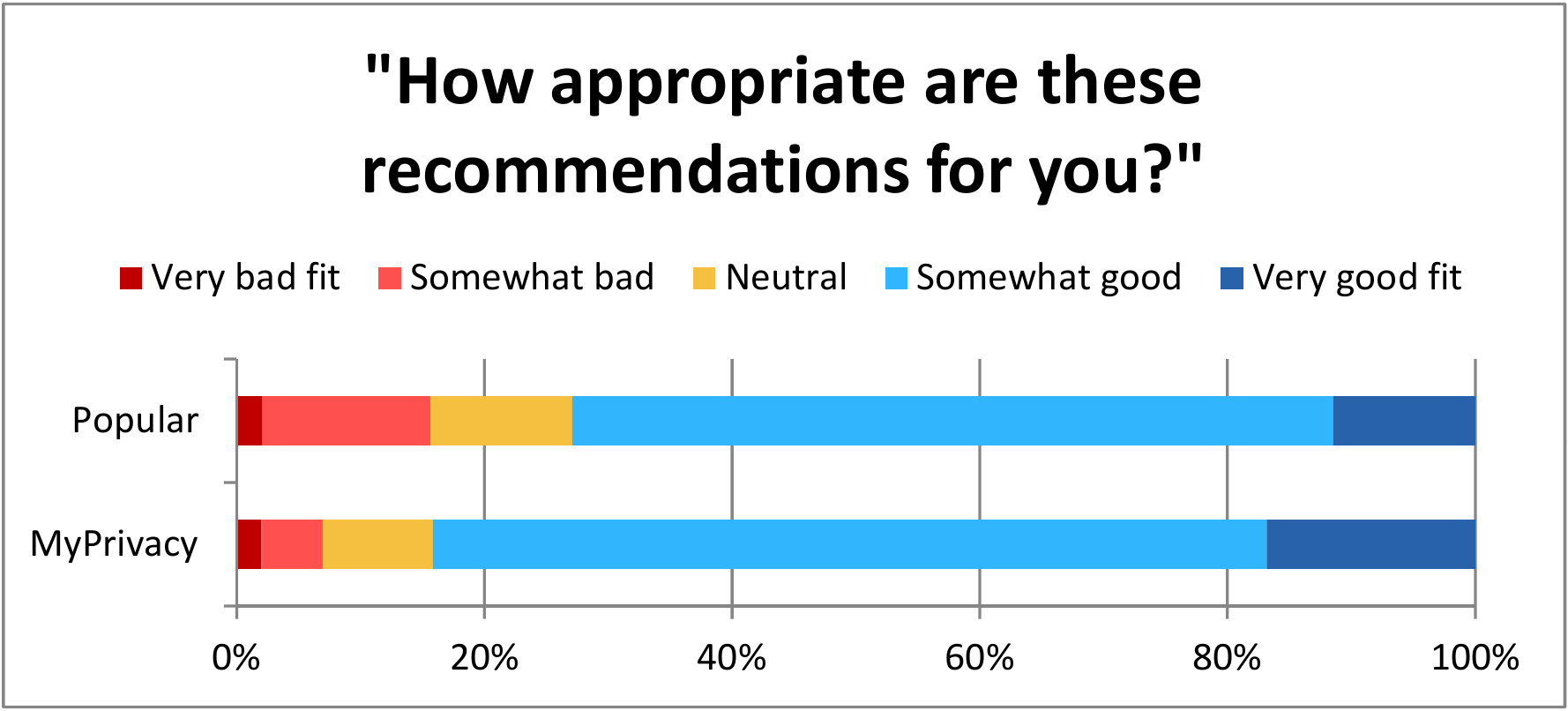}\label{fig:better}}\hspace{2em}
 \subfigure[Privacy of settings]{\includegraphics[width=\columnwidth]{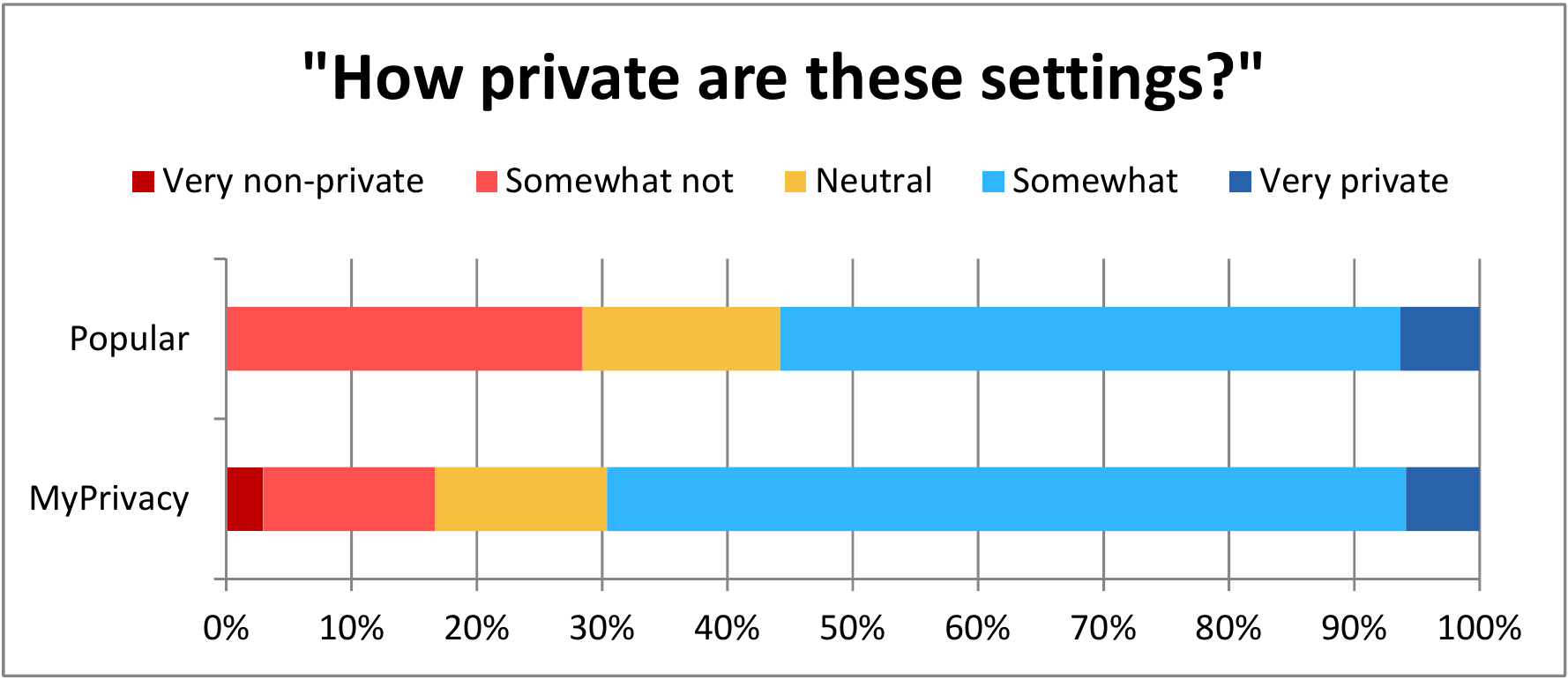}\label{fig:moreprivate}}
 \subfigure[Intent to apply recommendations]{\includegraphics[width=\columnwidth]{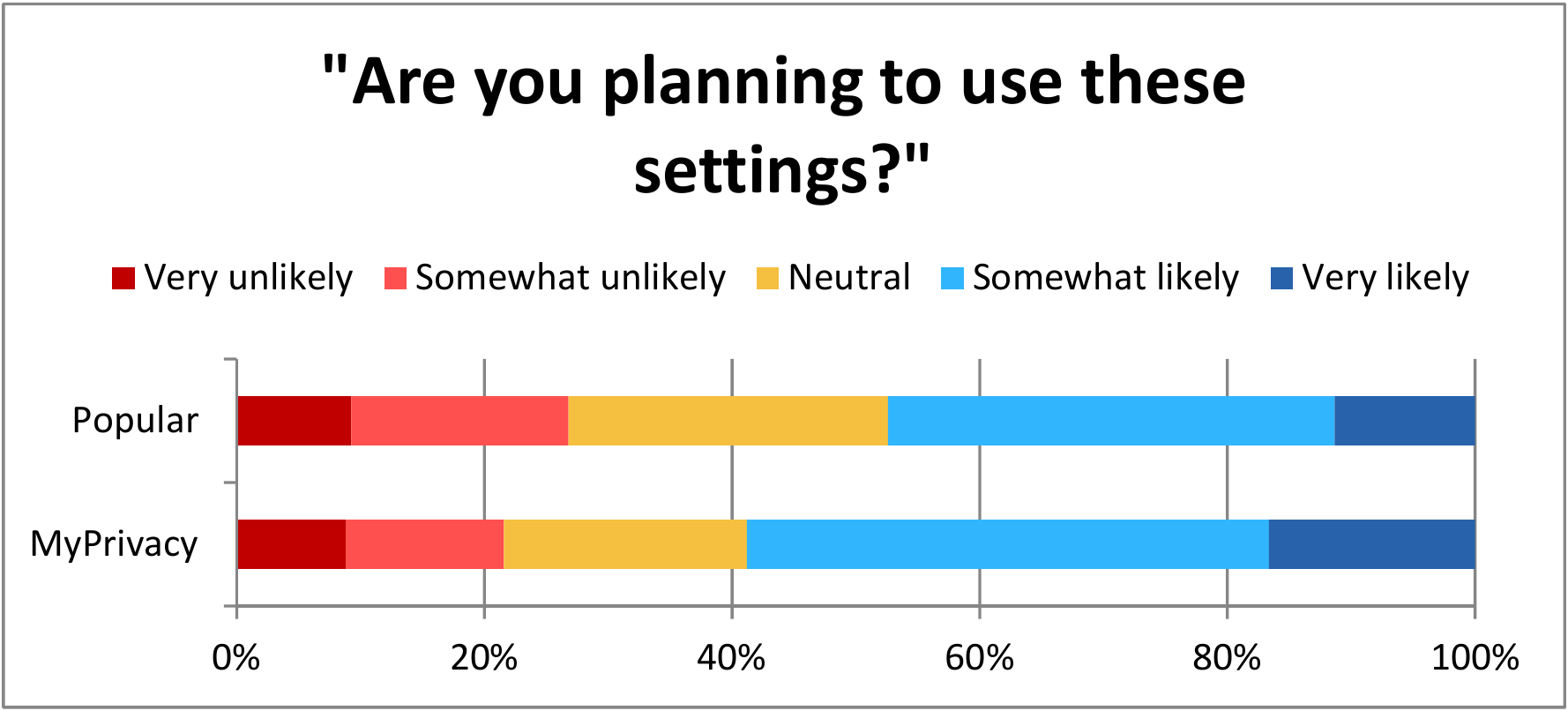}\label{fig:planuse}}\hspace{2em}
 \subfigure[Preference to use system]{\includegraphics[width=\columnwidth]{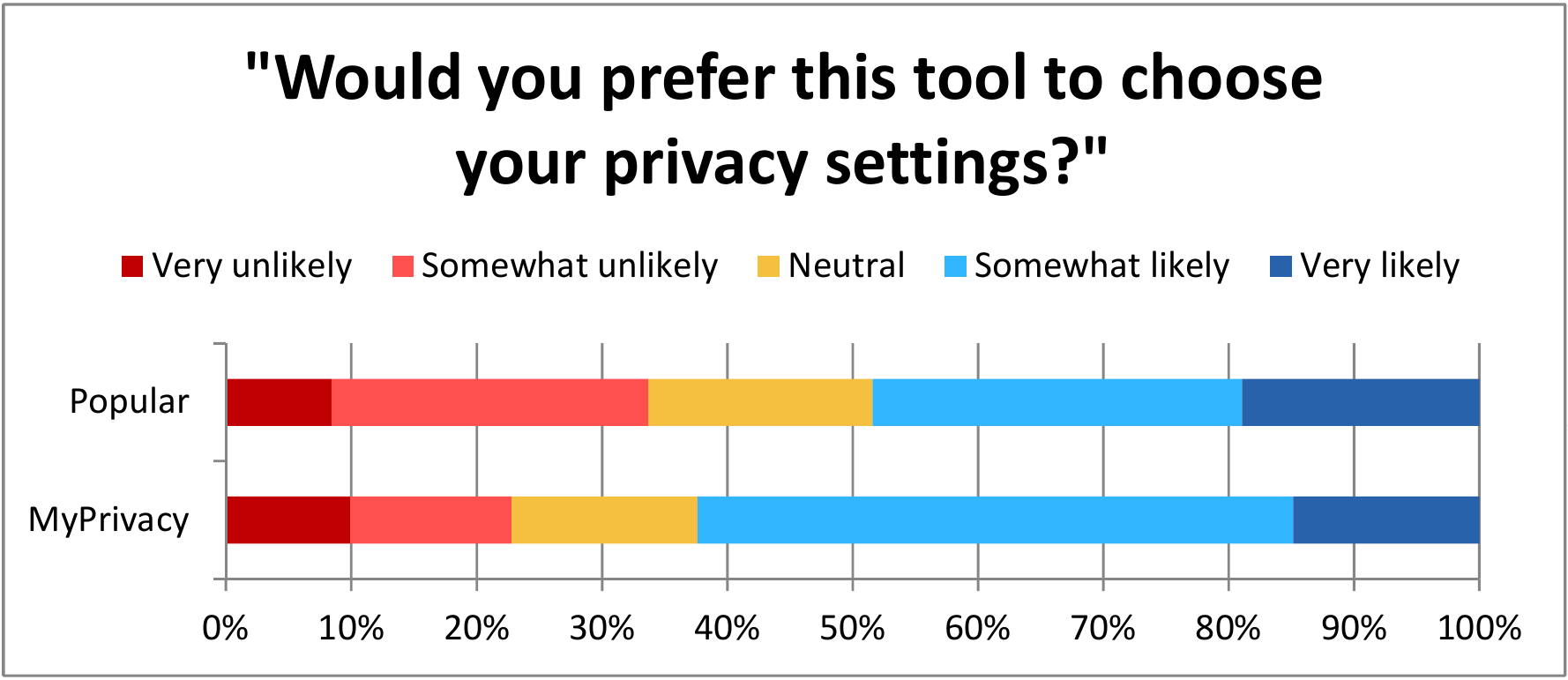}\label{fig:preferuse}}
 \caption{Users' feedback regarding different metrics comparing MyPrivacy against the most popular Facebook settings}\label{fig:eval}
\end{figure*}

As Figure \ref{fig:eval} shows, users perceived a clear distinction between the two models despite their identical representation as personalized recommenders. MyPrivacy outperformed the static popular recommendations in all four measures: appropriateness of the settings, privacy of the settings, users' intent to apply the changes, and users' preference to utilize the system in future interactions.

Specifically, with regard to the appropriateness of the recommendations, 84\% of the users who had tested MyPrivacy found its recommendations to be somewhat or very appropriate; only 7\% thought they were a very bad or somewhat bad fit. In contrast, 
the popular settings were considered appropriate by 73\% of users and inappropriate by 16\%.

This breakdown of opinions was mirrored in the users' assessment of the three modes' privacy. In response to the question, ``How private are these settings?", 70\% of the MyPrivacy users thought they were somewhat or very private. Among the users who tested the popular mode, 56\% considered the settings private. 

As compared with the baseline, users were more likely to implement the recommendations made by MyPrivacy. 
59\% of the users who tested MyPrivacy stated that they were somewhat or very likely to use the recommended settings, as compared to 47\% of the users who tested the popular mode.

Likewise, when asked if they would prefer to use such a tool to choose privacy settings, user responses favored MyPrivacy by a wide margin. Of the users who surveyed MyPrivacy, 62\% were somewhat or very likely to prefer such a tool; only 23\% said they were somewhat or very unlikely to prefer it. For the popular settings, 48\% were likely to use this tool and 34\% were unlikely. 

\subsubsection{Qualitative Responses} 

The recommendations of the MyPrivacy test mode met with a generally positive reception in the comment field. One user commented, ``This is a perfect match for me!" Another said, ``The recommendations are appropriate and I am more than likely to use it.'' A third wrote, ``This tool is really impressive. Thanks.'' A fourth user was very eager to implement the changes, asking, ``How can I get this information again so that I can change my privacy setting according to this settings?''

However, a few users expressed skepticism; for example, one wrote, ``Recommendations seem completely arbitrary -- how can you know my privacy needs based on some silly questions?'' Another said, ``I feel like I should have been asked more assessment questions before recommendations were made.'' Other users expressed a difficulty with mentally modelling the system, writing comments such as, ``I'm not sure where the suggestions came from  but they're very close to the privacy settings I use already. I'd still rather set my own privacy settings manually.'' This points to some ambivalence on the users' part: while overall they appreciate they utility of the system, the application would benefit from a better explanation of its underlying mechanisms to increase trust in its recommendations.

 With regard to the user interface, subjects appreciated how the color scale made the privacy trade-offs clearer in a given configuration of privacy settings. One user wrote, ``The color coding simplified things.'' Another commented, ``I like the color system of the website. It shows areas this person would need work on.''

Overall, the reconfiguration of Facebook privacy settings as a personalized recommender system seemed to strike users positively. One user commented, ``It seems much more convenient than having to manually input the options on my own, and if any differences do occur I can simply change one or two settings as opposed to changing all of them. I think it's a clever and useful idea.'' Likewise, another user wrote, ``I like the idea of a program that customly chooses your privacy settings.'' This is in line with the idea that Facebook users desire an alternative allowing them to configure their privacy settings in a more straightforward, uncomplicated way.

We did not specifically target people who had retained the default Facebook settings; however, we believe that such people would find MyPrivacy particularly useful as a way to guide them through the process of selecting the ``right'' settings. By projecting the difficult questions of privacy settings onto a more personal dimension, MyPrivacy makes good privacy decisions accessible to everyone. 
\section{Discussion and Future Work}
\label{sec:discussion}

Privacy facilitates trust, and trust encourages users to engage socially online. However, the state of the art in privacy lags far behind the times. In most services, users' preferences are defaulted to a one-size-fits-all privacy configuration, which they can then tweak at their prerogative. However, this assumption is flawed. Users express wildly different privacy preferences, and personalizing their options makes the task of privacy management much easier. Users are diverse in ages, cultures, and personalities, and their needs would be better served if this information were incorporated into the initial settings applied to their accounts. When we examined Facebook as a case study, we found clear support for this assertion in the fact that privacy settings varied according to certain personality and demographic traits.

Personalization is a much-vaunted tool for increasing content relevance across the Web, and we propose that it be used to increase
the relevance of a user's privacy defaults. We suggest that this be applied to Facebook's privacy settings. Since Facebook already has access to a massive amount of
data about its users, it would not be a difficult task to learn what settings different types of
users prefer. These preferences could then be offered as personalized defaults for new users,
updating the outmoded ``one size fits all'' paradigm of defaults on Facebook.

We believe that a more customized and helpful approach to privacy on Facebook's part
would increase trust in the Facebook privacy model, thus allowing users to feel more safe
about sharing information with their network and ultimately enhancing the viability of a Facebook business model.

\subsection{Future Work}
In the future, we could expand this work in a number of ways. Firstly, we could deploy similar analysis techniques on other social media platforms, such as LinkedIn and Twitter; the results would arguably be very different, due to the disparate functions that they serve in society. Additionally, our findings encourage more research on demographic factors, perhaps such as income or occupation. More examination of these may provide useful insights into what influences users' privacy decisions.

It is also important to consider other potential applications of this idea. In privacy and security applications, users are too often left to their own inadequate devices. However, simple heuristics and background data could be employed to make the process less laborious and its outcome more secure. For example, a browser has many components that can leave a user open to privacy and security breaches. Cookies enable surveillance, and password saving can be dangerous in case of a breach. Untrusted plugins can also do much damage. However, in many browsers, these features are turned on by default. If a browser could automatically infer a user's likely preferences, it would go a long way towards securing internet users and their data. 

In the social arena, this could also be extended to photo sharing applications. Certain types of users are more likely to want to share their images, a fact which could be deduced and applied to the default settings with some straightforward machine learning techniques. Likewise, location-based services could apply these ideas towards giving new users a more appropriate start to their privacy management by incorporating personalized privacy defaults. 
\section{Conclusion}
\label{sec:conclusion}
Can personalization function as tool to provide better privacy controls in online communities and services? In this paper, we proposed and analyzed the application of personalization based on individual factors to a user's privacy preferences. As a case study, we investigate users' privacy choices on Facebook. We found that neuroticism, age, ethnicity, and concern for privacy were related to the privacy of a user's Facebook configuration. As an application of our findings, we implement MyPrivacy, a personalized recommender system for Facebook privacy settings. Its evaluation shows that Facebook users are eager to welcome more straightforward efforts towards more personalized settings. Beyond that, it opens new vistas in the realm of usable privacy and security for online services and communities.



\bibliographystyle{abbrv}
\bibliography{sigproc}

\end{document}